\documentclass[preprint,aps]{revtex4}
\usepackage{graphicx}
\usepackage{amssymb}
\usepackage{amsmath}
\usepackage{epsfig}
\usepackage{stmaryrd}

\begin{document}

\title{Self-detecting gate-tunable nanotube paddle resonators}
\author {B. Witkamp, M. Poot, H. Pathangi, A. K. H\"{u}ttel and H. S. J. van der Zant\footnote{h.s.j.vanderzant@tudelft.nl}}
\affiliation{Kavli Institute of Nanoscience, Delft University of Technology, Lorentzweg 1, 2628 CJ Delft, The Netherlands} \date{\today}
\vspace{1cm}
\begin{abstract}
We have fabricated suspended metal paddle resonators with carbon nanotubes functioning as self-detecting torsional springs. We observe gate-tunable resonances, that either tune to higher or to lower frequencies when increasing the dc voltage on the back-gate. We attribute the former modes to flexural vibrations of the paddle resonator, while the latter ones are identified as torsional vibrations. Compared to top-down silicon fabricated paddle resonators, nanotube springs have smaller torsional spring constants and provide a larger frequency tunability.
\end{abstract}

\maketitle
\newpage
Mechanical resonators made from carbon nanotube (CNT) components show great promise in nano-electromechanical systems (NEMS). Due to their small radii, a deflection of only a few nm induces strain (tension) that significantly changes the properties of the resonators \cite{nature-Sazanova:284, nl-Witkamp:2904}. Nanotubes are furthermore ideal as building blocks for resonators due to their electronic properties, which can be modified by uniaxial or torsional strain \cite{prl-minot:156401, nnano-Cohen:36, nnano-hall:413} or by a gate-induced charge \cite{nature-Sazanova:284, nl-Witkamp:2904}. Recently, paddle resonators with nanotube springs have been fabricated and measured using transmission electron or optical microscopy \cite{science-Meyer:1539,nature-Fennimore:408,prl-papadakis:146101}. The small nanotube radii in these devices, lead to very small torsional spring constants ($10^{-15}-10^{-18}$ Nm/rad) so that they are easily twisted \cite{science-Meyer:1539, nnano-Cohen:36}, which makes them ideal for sensing applications. A paddle resonator is expected to have mechanical vibration modes with both flexural and/or torsional components. In this Letter, we present the first measurements on the dynamic behavior of paddle resonators with self-detecting nanotube springs and show that torsional and flexural modes can be distinguished based on their gate-tunability. This identification may lead to mode specific sensor applications.

\begin{figure}[tbp]
	\centering		
\includegraphics[width=16cm]{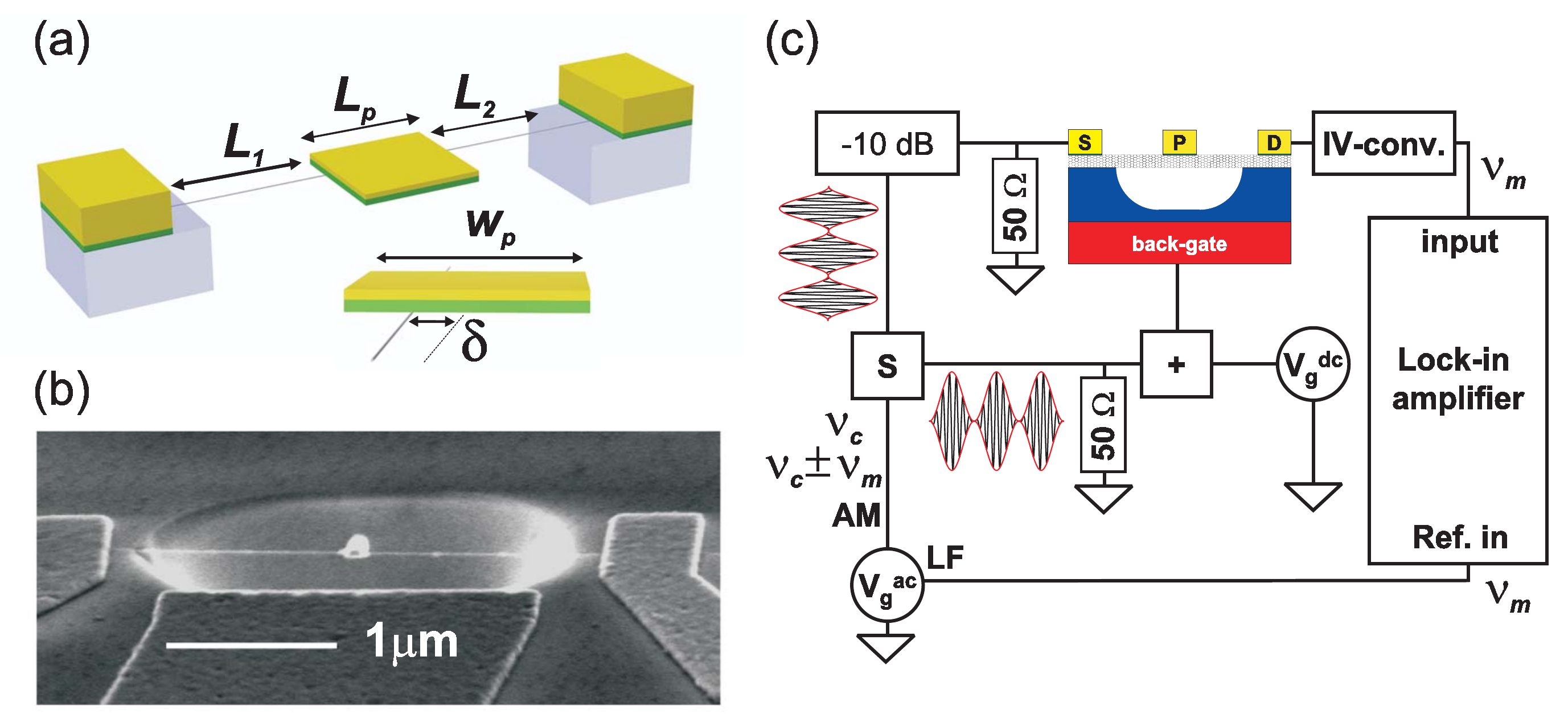}
	\caption{(a) Schematics of a paddle resonator defining some of its characteristic parameters. $\delta$ is the offset of the center of mass of the paddle from the nanotube axis, $L_{p}$ is the length of the paddle parallel to the nanotube, and $w_{p}$ is the the width of the paddle (perpendicular to the nanotube). (b) Scanning electron micrograph of a typical device. (c) AM frequency mixing measurement scheme. A single generator is used to generate an AM signal, which is split and supplied to the source (attenuated) and gate electrode. The modulation frequency is the reference for a lock-in amplifier. The current flowing through the nanotube is converted into a voltage and detected by the lock-in amplifier input.}
	\label{fig1}
\end{figure}

Devices (schematics of the device are shown in Fig. \ref{fig1} (a)) are made on a highly $p^{++}$ doped silicon wafers with 1 $\mu$m silicon oxide layer on top. The doped silicon is used as a back-gate electrode to actuate the paddle resonators electrostatically. Nanotubes are grown by means of chemical vapor deposition \cite{nt-babic:327} and located with respect to predefined reference markers. Source and drain electrodes (5 nm chromium and 50 nm gold) are defined with e-beam lithography, followed by a separate lithography step to define the metal paddles (10 nm chromium and 5-10 nm gold) on top of the nanotube. The samples are annealed at 400 $^\circ$C in argon to improve the adhesion of the paddle to the nanotube. To suspend the devices, an etch-mask is defined by e-beam lithography, followed by a wet-etch with a buffered HF solution. A scanning electron micrography (SEM) image of a typical paddle resonator is shown in Fig. \ref{fig1}(b).

The devices are mounted onto a custom made printed circuit board with an on-chip bias-T and 50 $\Omega$ terminator. The mounted sample is placed into a vacuum chamber, which is pumped down to a pressure of $10^{-5}$ mbar. The paddle resonator is electrostatically actuated with an amplitude modulated (AM) voltage $V_{g}^{ac}$ ($100\%$ modulation, carrier frequency $\nu_{c}$ and modulation frequency $\nu_{m}$) on the back-gate electrode (see Fig. \ref{fig1} (c)). A dc gate voltage $V_{g}^{dc}$ is used to statically deflect and/or rotate the paddle towards the back-gate \cite{nature-Sazanova:284, nl-Witkamp:2904, prl-hall:256102}. Measurements are performed at room temperature.

When the carrier frequency approaches a mechanical resonance, the nanotube and paddle oscillation amplitude increases significantly. Mechanical motion of the paddle has two effects; it changes the capacitance between the nanotube/paddle and the back-gate, and strain is induced (either torsional or longitudinal) in the CNT-springs. These effects give rise to capacitively induced \cite{nature-Sazanova:284,nl-Witkamp:2904} and/or piezoresistive \cite{nnano-Cohen:36,nnano-hall:413} changes of the conductance of the springs, and these changes are used to detect mechanical motion of the resonator. Since both mechanisms can occur simultaneously, a priori, no distinction can be made between the contributions. Both contributions to the ac conductance are described in more detail in the supplementary information.

To measure the mechanical conductance oscillations, which range between 1-20 MHz for our devices, we use the paddle resonators as frequency mixers \cite{nature-Sazanova:284,nl-Witkamp:2904}. This is done by biasing the paddle resonator with an AM-signal that is spectrally \textit{identical} to the voltage applied to the gate. The resulting current flowing through the device then has a spectral component at the modulation frequency $\nu_{m}$. This current is converted to a voltage and read-in with a lock-in amplifier using the modulation frequency (500 Hz) as the reference signal. The lock-in current is then:
\begin{eqnarray}
I^{LIA}=I^{ac}\cos(2\pi\nu_{m}t+\xi), \label{eq:liacurrent}
\end{eqnarray}
where the amplitude $I^{ac}$ and phase $\xi$ contain information about the electrostatic ($G^{ac}_{el}$) and mechanical ($G^{ac}_{mech}$) conductance changes (see supporting information for detailed analysis of the mode shapes etc.).

\begin{figure}[tbp]
	\centering		
   	\includegraphics[width=16cm]{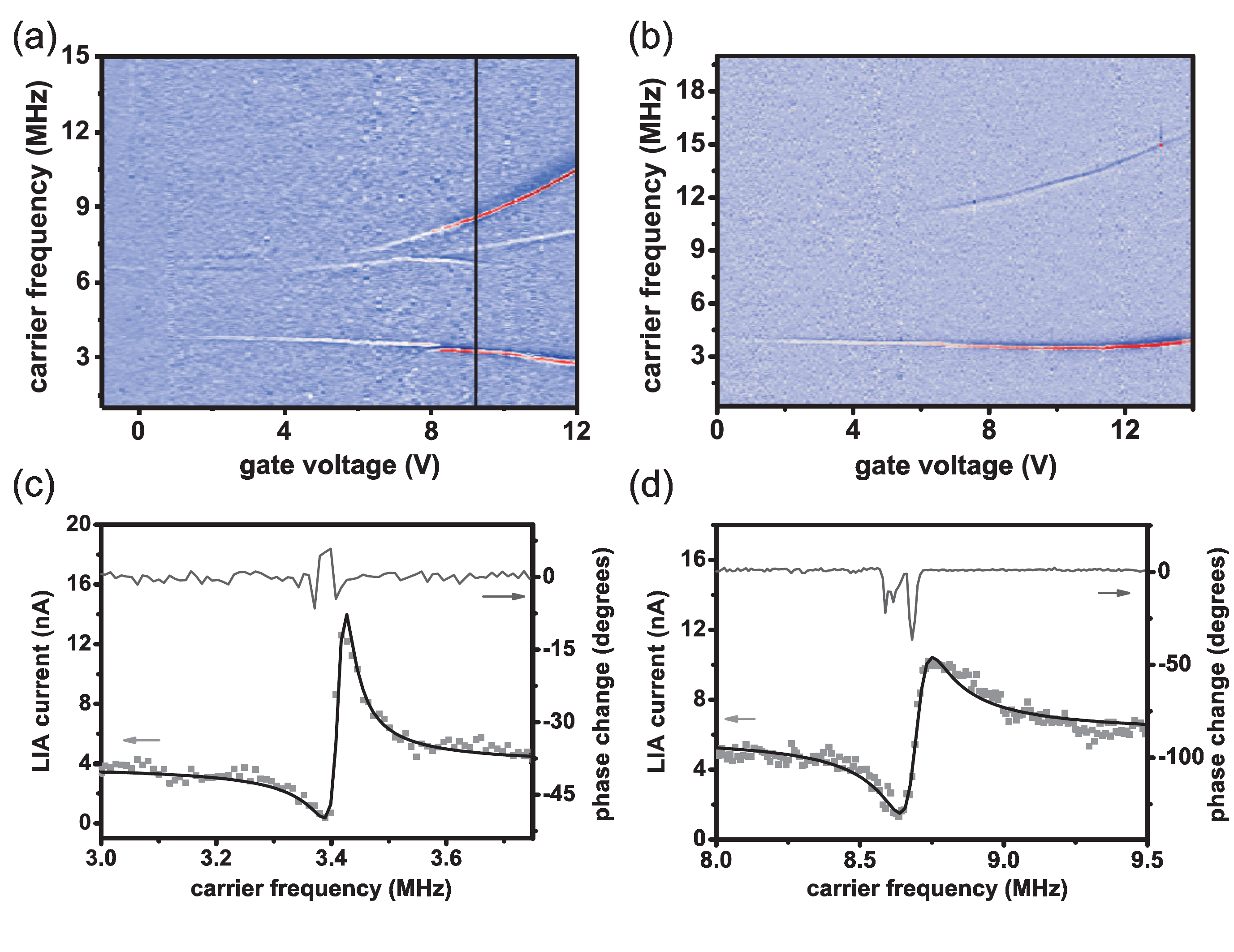}
	\caption{Mixing-current amplitude (differentiated with respect to $\nu_{c}$ to make the resonances clearly visible) as function of the carrier frequency vs dc gate voltage for (a) device 1 and (b) device 2 ($V_{g}^{ac}\approx$ 200 mV). Multiple resonances are visible that move either to higher or lower frequencies as the gate voltage is increased. (c) Peak shape of a typical down-tuned resonance measured at $V_{g}^{dc}=9.07$ V for device 1 (grey squares). The fit parameters are a resonance frequency of 3.41 MHz, $Q=214$ and a back-ground phase $\phi=0.16 \pi$ radians. The upper part depicts the relative phase vs carrier frequency. (d) Amplitude of a typical up-tuning resonance measured at $V_{g}^{dc}=9.07$ V for device 1. The fit parameters are a resonance frequency of 8.70 MHz, $Q=154$ and a back-ground phase $\phi=0.99\pi$ radians. Also for this mode a resonance is observed in the phase of the lock-in amplifier (upper gray curve).}
    \label{fig2}
\end{figure}

In more than 19 nanotube-paddle devices we have measured gate-tunable resonances in both the lock-in current and the phase, when sweeping the carrier frequency $\nu_{c}$. For two of these devices, with CNT radii of 1.4 nm and 2.2 nm for device 1 and device 2 respectively, Fig. \ref{fig2} (a) and (b) show the mixing-current amplitude as a function of both the carrier frequency and the dc gate-voltage applied to the back-gate electrode. We observe several modes that can be classified in resonances that are tuned to higher frequencies for increasing $V_{g}^{dc}$ or resonances that are tuned to lower frequencies. Note that in the devices that show resonances, we always observe the up-tuned resonances, while the down-tuned resonances are visible in just four devices. Typical lock-in amplitude and phase line traces are shown for a down-tuned resonance in Fig. \ref{fig2}(c) and an up-tuned resonance in Fig. \ref{fig2} (d). Black lines are fits of the amplitude $I^{ac}$ (Eq. \ref{eq:liacurrent}) to the data, and give information about the quality factor $Q$. In general, down-tuned modes have $Q$-factors ranging between 150-250, while up-tuned modes have $Q$'s ranging between 50-160.

In order to associate the tunability of observed resonances to specific modes (flexural or torsional vibrations), the torque and force acting on the paddle need to be determined. Due to the fringing fields the forces and torques acting on the paddle are difficult to express in an analytic form. Finite element (FEM) simulations of the capacitance $C_{g}$ between the paddle and the back-gate as a function of torsional angle $\theta$ show that for values of $\delta$ $>10$ nm (see Fig \ref{fig1} (a)) and angles up to 45 degrees, the capacitance ($C_{g}$) can be accurately described by a third order polynomial in $\theta$ (with coefficients $B_{0}$, $B_{1}$ and $B_{2}$ that need to be determined for each geometry). The torque is then described by a second order polynomial:
\begin{eqnarray}
T_{\theta}&=&\frac{1}{2}\frac{\partial C_{g}}{\partial \theta}V_{g}^{2}\approx \frac{1}{2} (B_{1} + 2 B_{2} \theta )V_{g}^{2}.\label{eq:torq}
\end{eqnarray}
The electrostatic force acting on the paddle can be approximated by:
\begin{eqnarray}
F&=&\frac{1}{2}\frac{\partial C_{g}}{\partial y}V_{g}^{2}\approx \frac{1}{2}(\frac{\partial B_{0}}{\partial y}+\frac{\partial B_{1}}{\partial y}\theta+\frac{\partial B_{2}}{\partial y}\theta^{2}   )V_{g}^{2}.\label{eq:force}
\end{eqnarray}
Note that the force and the torques are coupled to each other via the torsional angle. The dc and ac forces/torques are found by substituting $(V_{g})^{2}=(V_{g}^{dc})^{2}+2V_{g}^{dc}V_{g}^{ac}+(V_{g}^{ac})^{2}$. For our device geometry, the derivatives of coefficient $B_{1}$ and $B_{2}$ with respect to the displacement (Eq. \ref{eq:force}) are smaller than the derivative of coefficient $B_{0}$ by almost two orders of magnitude, so that the dc force acting on the paddle is dominated by the coefficient $B_{0}$.

For the devices used in our experiments the mass of the nanotube and the moment of inertia of the nanotube are at least two orders smaller than that of the Cr/Au-paddles (for both devices the paddle is made of 10 nm Cr and 10 nm Au), so that the mass of the CNT and its moment of inertia can be neglected. The ac displacement of the paddle can then be found by treating the paddle as a driven damped harmonic oscillator \cite{book:Cleland}, with flexural spring constant $\kappa$, driven by force $F$ (Eq. \ref{eq:force}). The spring constant $\kappa$ is the sum of the spring constants of the individual CNT sections.

To describe the flexural spring constants, we develop an adapted Euler-Bernoulli continuum model (see supplementary information) of the nanotube between the clamping point and the paddle island. The spring constant of an individual CNT section with length $L_{i}$ (where $i=1,2$) under influence of a force $F$ acting at the nanotube-paddle interface, is $12EI/L_{i}^{3}$ for $V_{g}^{dc}=0$ V. Here $I=\pi r^{4}/4$ is the second moment of inertia  of the nanotube with radius $r$, and Young's modulus $E$. For non-zero gate voltages \cite{pssb-poot:4252}, the spring constant has a $(V_{g}^{dc})^{4}$-dependence in the weak bending limit, and a $(V_{g}^{dc})^{4/3}$-dependence in the strong bending regime. Thus, the spring constant of the flexural modes increases in magnitude with gate-voltage, so that the resonance is tuned to higher frequencies.

The torsional vibration mode frequency $\nu_{t}$ of the paddle can be found by the balance of angular momentum \cite{book:Cleland}:
\begin{eqnarray}
I_{p}\ddot{\theta}+\frac{2I_{p}\nu_{t}}{Q}\dot{\theta}+k\theta=T_{\theta}, \label{eq:torqbal}
\end{eqnarray}
where $I_{p}$ is the moment of inertia of the paddle around the nanotube axis. In equation \ref{eq:torqbal} the total dc torsional spring constant is $k= I G(1/L_{1}+1/L_{2})$ \cite{comment:strain}, with $G=0.41$ GPa the shear modulus of the nanotube \cite{nnano-hall:413}. Since the torque in Eq. \ref{eq:torq} contains a $\theta$-dependence, and FEM simulations show that for our devices the coefficient $B_{2}$ is always positive, the effective torsional spring constant decreases with increasing gate-voltage. As a consequence, the resonance frequency \cite{joap-evoy:6072}
\begin{eqnarray}
\nu_{t}=\frac{1}{2\pi}\sqrt{\frac{k-\frac{\partial T_{\theta}}{\partial\theta}}{I_{p}}}\approx\frac{1}{2\pi}\sqrt{\frac{k-B_{2} (V_{g}^{dc})^{2}}{I_{p}}}\label{eq:torsfreq}
\end{eqnarray}
\textit{decreases} with increasing $V_{g}^{dc}$, in contrast to flexural modes, which are tuned to higher frequencies. Note that previous work on silicon paddle resonator devices \cite{joap-evoy:6072} and capacitively tunable beam resonators \cite{apl-kozinsky:253101} have also shown a down shift with gate voltage, but much less pronounced. This makes the nanotube based device particularly suitable for applications that require highly tunable resonators.

\begin{figure}[tbp]
	\centering		
   	\includegraphics[width=16cm]{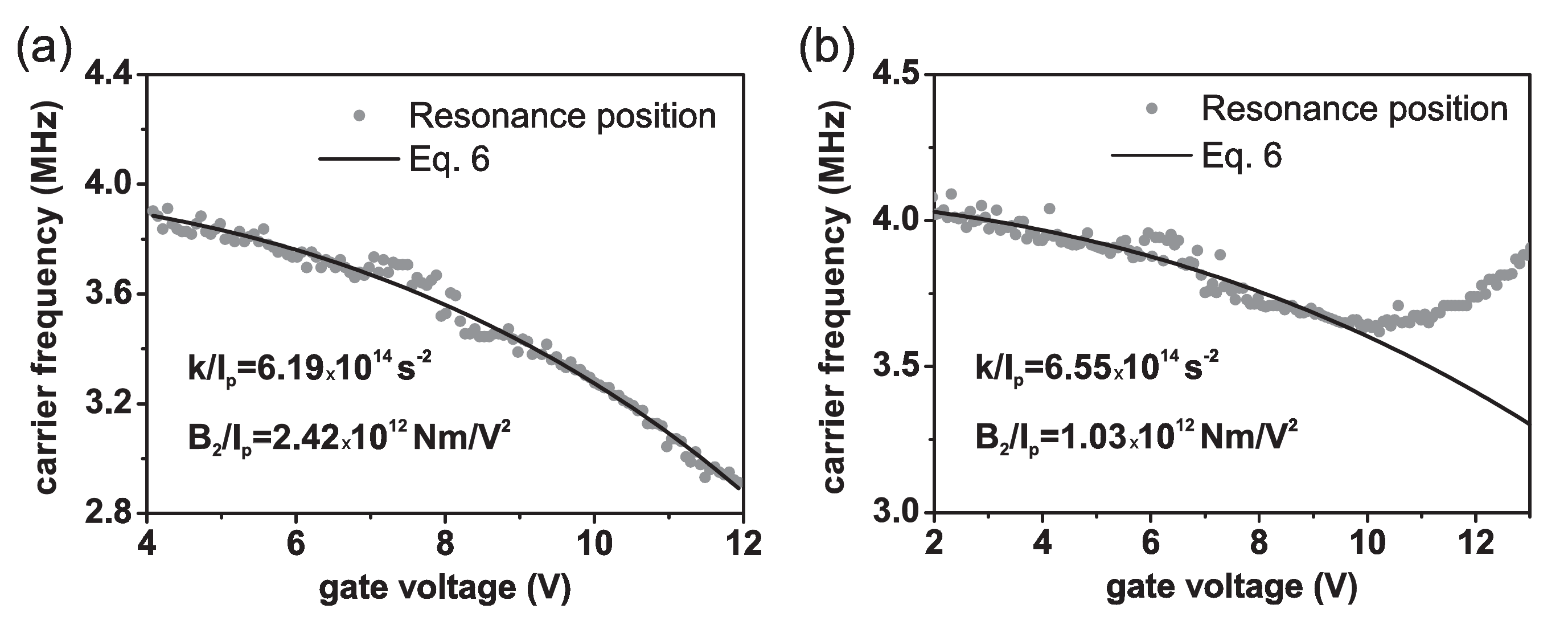}
	\caption{
Resonance positions (points) extracted from the lock-in current as a function of carrier frequency vs dc gate voltage for device 1 (a) and 2 (b). The solid lines are fits of Eq. \ref{eq:torsfreq} to the datapoints in which $k/I_{p}$ and $B_{2}/I_{p}$ are the fit parameters. }\label{fig3}
\end{figure}

A more quantitative analysis of the down-tuned torsional modes can be made from the measurements of the lock-in current (Fig. \ref{fig2} (a) and (b)) as a function of gate-voltage and driving frequency. We have extracted the peak positions (black points) of the down-tuned resonance of device 1 and 2 (Fig. \ref{fig3}(a) and (b)). By fitting Eq. \ref{eq:torsfreq} to the gate dependence of the resonance frequency and calculating the moment of inertia $I_{p}$ (for device parameters see supplementary information), we find $B_{2}=8.3 \times 10^{-20}$ Nm/V$^{2}$ for device 1 and $B_{2}=3.3 \times 10^{-20}$ Nm/V$^{2}$ for device 2. From FEM calculations we find $B_{2}=3.6 \times 10^{-21}$ Nm/V$^{2}$ and $B_{2}=7.0 \times 10^{-21}$ Nm/V$^{2}$ respectively, which is in agreement with the experimental values, considering the uncertainty in the nanotube radii and the thickness of the paddle.

We have also calculated the gate-dependence of the flexural modes numerically using the model described above. We find that the gate dependence of the flexural mode is in qualitative agreement with the up-tuned resonances, but that the magnitude is too low. The difference in magnitude can be explained by a non-zero force acting on the nanotube sections. Our model, does not take into account asymmetric spring constants, which would give rise to more complicated vibration modes that contain both flexural and torsional components. This would explain the change in tunability of the resonance in Fig. \ref{fig3} (b) for $V_{g}^{dc}$ $>$ 10 V; at high gate voltages the flexural component of the vibration mode becomes larger than the torsional part. A more sophisticated model may be required to describe these resonances.

In conclusion, we have observed mechanical resonances of self-detecting metal paddle resonators with CNT-springs. We also demonstrate that, torsional modes and flexural modes can be distinguished from each other based on the gate-tunability of these modes. This distinct tuning behavior provides a tool to study for example mass or force sensitivity of different vibration modes.

Financial support is obtained from the Dutch organizations FOM, NWO (VICI-grant) and NanoNed.

\newpage

\newpage

\newpage

\newpage


\begin{thebibliography}{99}
\bibitem{comment:strain}
The effects of (gate tunable) longitudinal strain $\epsilon$ on the nanotube lengths are neglected; longitudinal strain increases the length of each nanotube section and thereby decreases the torsional spring constants, but the effect is too small for the parameters in our study (strain is of the order of $10^{-6}$). From the strain it is therefore a priori not clear if torsional modes tune with gate voltage.
\end{thebibliography}
\end{document}